\documentclass{ws-procs9x6}

\usepackage{latexsym}

\newcommand{\be}{\begin{equation}} \newcommand{\ee}{\end{equation}}
\newcommand{\bea}{\begin{eqnarray}} \newcommand{\eea}{\end{eqnarray}}

\def\am{Center For Theoretical Physics, Department of Physics\\ Texas A\&M University,
 College Station,TX 77843-4242,
USA}
\def\address#1{\begin{center}{ \it #1} \end{center}}
\def\author#1{\begin{center}{ \sc #1} \end{center}}
\def\title#1{\begin{center} {\Large #1 } \end{center}}
\def\Journal#1#2#3#4{{#1} {\bf #2}, #3 (#4)}

\def\NPB{{\em Nucl. Phys.} B}

\def\PRL{\em Phys. Rev. Lett.}
\def\PRD{{\em Phys. Rev.} D}

\begin{document}


\title{Gravitational Forces on the Branes}
\author{R. Arnowitt and J. Dent}
\address{\am}

\abstracts{\it{It is a great pleasure to present this work in honor of Pran Nath, whose many important contributions to quantum field theory, supergravity grand unification, and their consequences have been on the forefront of modern particle physics.}}
\abstracts{We examine the gravitational forces in a brane-world scenario felt by point particles on two 3-branes bounding a 5-dimensional AdS space with $S^{1}$/$Z_2$ symmetry. The particles are treated as perturbations on the vacuum metric and coordinate conditions are chosen so that no brane bending effects occur.  We make an ADM type decomposition of the metric tensor and solve Einstein's equations to linear order in the static limit.  While no stabilization mechanism is assumed, all the 5D Einstein equations are solved and are seen to have a consistent solution.  We find that Newton's law is reproduced on the Planck brane at the origin while particles on the TeV brane a distance $y_2$ from the origin experience an attractive force that has a growing exponential dependence on the brane position.}


\section{Introduction}
 Models of the universe consisting of a five-dimensional bulk space bounded by branes have been the subject of a great deal of study over the last several years.  The theoretical motivation for such work can be found in the Horava-Witten M-Theory (HW) \cite{hw,hw2,witten,horava}, which is the strong coupling limit of the $E_8 x E_8$ heterotic string theory.  In HW, eleven dimensional supergravity resides in the bulk that is bounded by two ten-dimensional planes, each of which is located at a fixed point of an $S^1/Z_2$ orbifold and carries one of the $E_8$ gauge groups.  Upon compactification of six dimensions on a Calabi-Yau threefold there is an intermediate energy range where the universe will appear five dimensional.  This is due to the fact that the interval separating the branes is $\mathcal{O}$(10) times larger than the compactification scale\cite{witten}.  The phenomenological studies of five-dimensional models were initiated by Binetruy et.al.\cite{binetruy} as well as Randall and Sundrum (RS)\cite{rs}.  In these and subsequent studies it was shown that one can reproduce the FRW cosmology at late times as well as using the location of the brane upon which our universe resides in the five-dimensional geometry as a possible solution to the hierarchy problem.  

In this paper we would like to determine the gravitational forces felt by point particles placed on the branes.  We will find that we will be able to recover the standard Newton's law only on the brane at the origin and that particles on the distant brane will feel a large attractive force which scales exponentially with its distance from the origin.    

We consider a 5D AdS bulk space bounded by two 4D orbifold planes.  This could be a 5D reduction of Horava-Witten theory with $S^1/Z_2$ symmetry, or a two brane Randall-Sundrum model \cite{rs}.  We will study the specific case where the 4D planes have vanishing cosmological constant.  The solution of the vacuum equations has been given in \cite{rs} as
\begin{equation}
ds^2 = e^{-2 A(y)}\eta_{ij}dx^{i}dx^{j} + dy^2\,\,\,;\,\,\, A(y)=\beta |y|
\end{equation}
Point particles are placed on the branes and we ask for the gravitational forces between them.  There is a large literature on this subject\cite{lykken,csaki,tanaka,giddings,chung,dorca,deruelle,callin}, on the question of recovering Newton's 4D law and the corrections to it arising from the presence of an additional dimension.  

We assume that matter is added on the branes as a perturbation to the vacuum metric
\begin{equation}
ds^2 = e^{-2\beta y}\left(\eta_{ij} + h_{ij}\right)dx^{i}dx^{j} + h_{i5}dydx^i + \left(1 + h_{55}\right)dy^2
\end{equation}
and then solve the Einstein equations to linear order in $h_{\mu\nu}$ where $\mu$ = 0,1,2,3,5, i = 0,1,2,3, and y = $x^5$.

The diffeomorphisms of a 5D theory with $S^1/Z_2$ symmetry are those of $R_{4}xS^1$ which commute with $Z_2$.  This means that for the transformation
\begin{equation}
x^{\mu} \rightarrow x^{\mu} + \xi^{\mu} \equiv x'^{\mu}
\end{equation}
one has that $\xi^{5}$ vanishes at the orbifold points, $y_1$ and $y_2$. 
\begin{equation}
\xi^{5}(x^0,y_1) = 0 = \xi_{5}(x^i,y_2)
\end{equation}
If one were to make a coordinate transformation with a non-vanishing $\xi^{5}$, then the branes become bent and this would create a complication when one imposes the $Z_2$ boundary constraint on the brane leading to so-called brane bending effects.

In previous analyses, the 5D Einstein equations are solved in Gaussian normal coordinates which are described by
\begin{eqnarray}
h_{5\mu} = 0\\
\partial^{j}h_{ij} = 0 = \eta^{ij}h_{ij}
\end{eqnarray}
In general, these cannot be achieved without brane bending occurring.  We give here an alternate analysis which avoids these complications.

\section{\it{Coordinate Conditions}}
In general, any symmetric array $h_{ij} = h_{ji}$ can be decomposed according to the ADM prescription\footnote{This decomposition was first introduced in \cite{adm}. (Ref.\cite{adm2} is a more accessible recent reprint summarizing the ADM formalism.)  The generalization to 4-space is trivial except for the ambiguity in defining 1/$\Box^2$ in Minkowski space.  However, we will always be considering the static Newtonian limit here where $\Box^2$ $\rightarrow$ $\nabla^2$, though the size (and correct definition of) the higher order dynamical effect are also of interest.} ($h_{i,j}$ $\equiv$ $\partial_{j}h_i$)
\begin{equation}
h_{ij} = h_{ij}^{TT} + h_{ij}^{T} + h_{i,j} + h_{j,i}
\end{equation}
where $h_{ij}^{TT}$ is transverse and traceless, $h_{ij}^T$ is transverse and can possess a trace, 
\begin{eqnarray}
\partial^{i}h_{ij}^{TT} \equiv 0 \equiv \eta^{ij}h_{ij}^{TT}\\
\partial^{i}h_{ij}^{T} \equiv 0\,\,\,;\,\,\, \eta^{ij}h_{ij}^{T} \equiv f^{T} \neq 0
\end{eqnarray}
and 
\begin{equation}
h_i = h_{i}^{T} + \frac{1}{2}h^{L}_{,i} \,\,\,;\,\,\, \partial^{i}h_{i}^{T} \equiv 0
\end{equation}
One can write
\begin{equation}
h_{ij}^{T} = \frac{1}{3}\pi_{ij}f^T \,\,\,;\,\,\, \pi_{ij} \equiv \eta_{ij} - O_{ij}
\end{equation}
where
\begin{equation}
O_{ij} \equiv \frac{\partial_i\partial_j}{\Box^2}
\end{equation}
One can express each of these subpieces in terms of $h_{ij}$, i.e.
\begin{eqnarray}
f^T = \pi^{ij}h_{ij} \\
h_{ij}^{TT} = \pi_{ik}\pi_{jl}h^{kl} - \frac{1}{3}\pi_{ij}\pi_{kl}h^{kl}\\
h^{L}_{,ij} = O_{ik}O_{jl}h^{kl}\\
h^{T}_{i,j} = O_{j}^{k}h_{ik} - O_{ik}O_{jl}h^{kl}
\end{eqnarray}

In order to simplify the anaysis one can choose the coordinate condition
\begin{equation}
h_{5i} = 0 
\end{equation}
Under a linearized coordinate transformation
\begin{equation}
x^{\mu} \rightarrow x^{\mu} + \xi^{\mu}(x;y)
\end{equation}
one can always achieve $h_{5i}$ = 0 using only the $\xi^{i}$ maintaining Eq.(4).  In terms of $\omega^{\mu}$ = $e^{2A(y)}\xi^{u}(x;y)$ the remaining coordinate freedom is then
\begin{eqnarray}
\omega_{i}^{T}(x;y) = F_{i}^{T}(x)\\
\omega_{5}(x;y) = -\omega^{L}_{,5}(x,y)\\ \omega_{5}(x;y_1) = 0 = \omega_{5}(x;y_2) 
\end{eqnarray}
where $\omega_{i}$ = $\omega_{i}^{T}$ + $\omega_{,i}^{L}$, $\omega^{L}_{,i} = O_{ij}\omega^{j}$, and $\omega^{T}_{i} = \omega_{i} - O_{ij}\omega^{j}$.  $F_{i}^{T}$ is an arbitrary function of x (independent of y).  This then allows a remaining gauge freedom on the metric of ($A'$ $\equiv$ $dA/dy$):
\begin{eqnarray}
\delta h_{55} = -2(e^{-2A}\omega^{L}_{5})_{,5}\\
\delta f^T = 6A'e^{-2A}\omega^{L}_{,5}\\
\delta(\Box^{2}h^L) = 2\Box^2\omega^L + 2A'e^{-2A}\omega^{L}_{,5}\\
\delta h^{T}_{i} = \omega^{T}_{i}(x)\,\,\,;\,\,\, \delta h_{ij}^{TT} = 0
\end{eqnarray}
Since by Eq.(4) we require $\omega_{5}(x;y_{\alpha})$ = 0, $f^{T}(y_{\alpha})$ is gauge invariant, and we can choose (using $\omega_5$)
\begin{equation}
\partial_{5}f^{T}(x,y_{\alpha}) = 0\,\,\,;\,\,\,\alpha = 1,2
\end{equation}

Note that since $h_{55}$ $\neq$ 0, the invariant distance between the branes may change, so the equations should still be consistent even though a Goldberger-Wise stabilization mechanism has yet to be included in the analysis.  However, a satisfactory RS model requires a stabilization mechanism and we will see what problems the lack of such a mechanism may produce.

\section{\it{Solution of the Einstein equations}}

The 5D Einstein equations read
\begin{equation}
R_{ij} = \frac{1}{M_{5}^3}(T_{ij} - \frac{1}{3}\eta_{ij}T) \,\,\, ; \,\,\, T = \eta^{mn}T_{mn}
\end{equation}
where $M_5$ is the 5D Planck mass.  Upon projecting out the transverse traceless modes, one is left with
\begin{equation}
[-\partial_{5}^{2} + 4A'\partial_5 - e^{2A}\Box^2]h_{ij}^{TT} = \frac{2e^{2A}}{M_{5}^3}\sum_{\alpha = 1,2}T_{ij}^{TT}\delta(y-y_{\alpha}). 
\end{equation}
subject to the $Z_2$ boundary conditions on the brane
\begin{equation}
\partial_{5}h_{ij}^{TT} = (-1)^{\alpha}\frac{e^{2A}}{M_{5}^3}T_{ij}^{TT}(y_{\alpha}) \,\,\,;\,\,\, \alpha = 1,2.
\end{equation}
We solve these equations by Fourier transforming the $x^i$ dependence 
\begin{equation}
h_{ij}^{TT}(x;y) = \int d^{4}p e^{ipx}h_{ij}^{TT}(p;y).
\end{equation}
For the static forces on point particles on the branes one needs (using Eq.7)
\begin{equation}
h_{00}(x,y_{\alpha}) = h_{00}^{TT}(x;y_{\alpha}) - \frac{1}{3}f^T(x,y_{\alpha}).
\end{equation}

On the Planck brane ($y_1$ = 0) we find
\begin{equation}
h_{00}^{TT}(p;y_1) = -\frac{2}{3\beta M_{5}^3}\left[\frac{N_{11}(\xi_1,\xi_2)}{D}T_{00}(y_1) + \frac{N_{12}(\xi_1,\xi_2)}{D}T_{00}(y_2)\right]
\end{equation}
where
\begin{eqnarray}
D \equiv \frac{N_1(\xi_1)}{J_1(\xi_1)} - \frac{N_1(\xi_2)}{J_1(\xi_2)} \\
\xi_1 = \frac{m}{\beta}\,\,\,; \,\,\,\xi_2 = \frac{m}{\beta}e^{\beta y_2}\\
m^2 \equiv -p^2 = (p^{0})^2 - \vec{p}^2\\
N_{11} \equiv \frac{J_2(\xi_1)}{\xi_1 J_1(\xi_1)}\left[\frac{N_2(\xi_1)}{J_2(\xi_1)} - \frac{N_1(\xi_2)}{J_1(\xi_2)}\right]\\
N_{12} \equiv \frac{J_2(\xi_1)}{\xi_2 J_1(\xi_2)}\left[\frac{N_2(\xi_1)}{J_2(\xi_1)} - \frac{N_1(\xi_1)}{J_1(\xi_1)}\right]
\end{eqnarray}
with $J_k$ and $N_k$ being the Bessel and Neumann functions of order k.  On the TeV brane ($y_2$ = $\pi\rho$) we find
\begin{equation}
h_{00}^{TT}(p;y_2) = -\frac{2e^{2\beta y_2}}{3\beta M_{5}^3}\left[\frac{N_{21}(\xi_1,\xi_2)}{D}T_{00}(y_1) + \frac{N_{22}(\xi_1,\xi_2)}{D}T_{00}(y_2)\right]
\end{equation}
where
\begin{equation}
N_{21}(\xi_1,\xi_2) = N_{12}(\xi_2,\xi_1)\,\,\,;\,\,\,N_{22}(\xi_1,\xi_2) = N_{11}(\xi_2,\xi_1)
\end{equation}

We use the other Einstein equations to determine the remaining components of the metric.  Thus for $R_{5i}$ one has
\begin{equation}
R_{5i} = -\partial_{i}\partial_{5}f^T - 3A'\partial_{i}h_{55} + \Box^2\partial_{5}h_{i}^{T} = 0
\end{equation}
The T part of Eq.(40) yields
\begin{equation}
\partial_{5}h_{i}^{T} = 0.
\end{equation}
Therefore $h_{i}^{T}$ is independent of y and can be set to zero by a gauge choice of $\xi_{i}^{T}$ (using Eq.(25) and the gauge freedom of Eq.(19)).  The remaining terms in Eqn.(40) then give $h_{55}$ in terms of $f^T$
\begin{equation}
h_{55} = \frac{1}{3A'}\partial_{5}f^T.
\end{equation}
The gauge choice of Eq.(26) then implies the vanishing of $h_{55}$ on the branes as well.  However, $h_{55}$ will not in general vanish in the bulk.

The two remaining variables, $h^L$ and $f^T$ can be obtained from the $R_{55}$ and $\eta^{ij}R_{ij}$ equations.  These equations read
\begin{eqnarray}
&&(\frac{1}{2}\partial_{5}^{2}-4A'\partial_5)(\Box^{2}h^L - \frac{1}{3}f^T) + e^{2A}\Box^{2}f^T - \frac{e^{2A}}{6A'}\Box^{2}\partial_{5}f^T\\\nonumber
&=& -\frac{e^{2A}}{3M_{5}^3}\sum_{\alpha = 1,2}T_{00}(y_{\alpha})\delta(y-y_{\alpha})
\end{eqnarray}
and
\begin{eqnarray}
&&(\frac{1}{2}\partial_{5}^{2}-A'\partial_5)(\Box^{2}h^L - \frac{1}{3}f^T) + \frac{4e^{2A}}{3}\Box^{2}f^T - \frac{e^{2A}}{6A'}\Box^{2}\partial_{5}f^T\\\nonumber
&=& -\frac{e^{2A}}{3M_{5}^3}\sum_{\alpha = 1,2}T_{00}(y_{\alpha})\delta(y-y_{\alpha})
\end{eqnarray}
with boundary conditions
\begin{equation}
\partial_{5}(\Box^{2}h^L - \frac{1}{3}f^T)\bigg|_{y=y_{\alpha}} = -\frac{(-1)^{\alpha + 1}e^{2A(y_{\alpha})}}{3M_{5}^3}T_{00}(y_{\alpha}).
\end{equation}
One can use these to eliminate $h^L$ in terms of $f^T$
\begin{equation}
\Box^{2}h^L = \frac{1}{3}f^T + \frac{1}{3}\int_{0}^{y}dy'\frac{e^{2A(y')}}{A'}\Box^{2}f^{T}(y')
\end{equation}
and the boundary conditions then determine
\begin{equation}
\Box^{2}f^{T}(x^i,y_{\alpha}) = (-1)^{\alpha}\frac{\beta}{M_{5}^3}T_{00}(y_{\alpha}).
\end{equation}
One should note that while $f^T$ is gauge variant in the bulk, it is gauge invariant on the branes (which is why its values on the branes are determined by the physical quantities $T_{00}(y_{\alpha})$).  We note also that the other Einstein equations contain no further constraints on the metric.

\section{\it{Poles of $h_{00}$($p^i$,$y_{\alpha}$)}}
In Fourier space
\begin{equation}
f^{T}(y_{\alpha}) = \frac{(-1)^{\alpha}\beta}{m^{2}M_{5}^3}T_{00}(y_{\alpha})\,\,\,;\,\,\,m^2 = -p^2
\end{equation}
which in the static limit gives a contribution to the 1/r Newtonian potential.  Note $f^T$ at $y_{\alpha}$ sees only $T_{00}(y_{\alpha})$, i.e., it does not see matter on the other brane.  On the other hand, $h_{00}^{TT}$ does see matter on both branes (Eqs.(32,38)) and therefore the total $h_{00}$ of Eq.(31) does as well.

The poles of $h_{00}^{TT}$ give contributions not only to the 1/r term but also give $1/r^2$, $1/r^3$, ..., corrections to the Newtonian potential (these corrections are from the KK modes). One can see that the 1/r term in $h_{00}^{TT}$ comes from the fact that as $m^2$ $\rightarrow$ 0,
\begin{equation}
N_2 \sim \frac{1}{\xi_{1}^2} \sim \frac{1}{m^2}
\end{equation}
in the $N_{ij}$ functions while the zeros of the denominator D give the higher order correction terms.  For example in the regime $\xi_1$$\ll$1 and $\xi_2$ $\gg$1 the poles occur at
\begin{equation}
\frac{m_n}{\beta} \cong [(n+\frac{5}{4})\pi + \frac{\pi^2}{16}(\frac{m_{n}^2}{\beta^2})^2]e^{-2\beta y_2}
\end{equation}
and for $\xi_1$$\gg$1 and $\xi_2$ $\gg$1 the poles are at
\begin{equation}
\frac{m_n}{\beta} \cong n\pi e^{-\beta y_2}.
\end{equation}
In each regime, for the respective cases of the RS model and HW theory, the poles begin at
\begin{eqnarray}
&(RS)&\,\,\,\, m_1 \approx \pi\beta e^{-\beta y_2} \approx (10^{19})(10^{-16}) = 1TeV \\
&(HW)&\,\,\,\, m_1 \approx 10^{15}e^{-\beta\pi\rho} \approx 10^{14}GeV
\end{eqnarray}
Here we have used the fact that in RS $e^{-\beta y_2}$ $\approx$ $10^{-16}$ to account for the gauge hierarchy.

In order to calculate the potential we consider $h_{00}^{TT}$($m^2;y_{\alpha}$) with complex variable $m^2$ and define
\begin{equation}
g(z) = \frac{h_{00}^{TT}(z;y_{\alpha})}{z-m^2}.
\end{equation}
Then by integrating g(z) over a large circle in the complex plane we can solve for $h_{00}^{TT}(m)$ as the sum (over $m_n$) of the residues of the poles of $h_{00}^{TT}$.
This results in
\begin{eqnarray}
&&h_{00}^{TT}(y_1) = -\frac{4\beta}{3m^{2}M_{5}^3}[T_{00}(y_1) + e^{-2\beta y_2}T_{00}(y_2)]\\\nonumber
&&+ \frac{4\beta}{M_{5}^3}\sum_{m_n}\frac{1}{m^2 - m_{n}^2}\left(\frac{J_{2}^{2}(\xi_2)}{J_{2}^{2}(\xi_2) - J_{1}^{2}(\xi_1)}\right)[T_{00}(y_1) + \frac{e^{-\beta y_2}J_{1}(\xi_1)}{J_{1}(\xi_2)}T_{00}(y_2)]\bigg|_{m=m_n}
\end{eqnarray}
and
\begin{eqnarray}
&&h_{00}^{TT}(y_2) = -\frac{4\beta}{3m^{2}M_{5}^3}[T_{00}(y_1) + e^{-2\beta y_2}T_{00}(y_2)]\\\nonumber
&&+ \frac{4\beta}{M_{5}^3}\sum_{m_n}\frac{1}{m^2 - m_{n}^2}\left(\frac{J_{1}^{2}(\xi_1)}{J_{1}^{2}(\xi_2) - J_{1}^{2}(\xi_1)}\right)[T_{00}(y_2) + \frac{e^{\beta y_2}J_{1}(\xi_2)}{J_{1}(\xi_1)}T_{00}(y_1)]\bigg|_{m=m_n}
\end{eqnarray}
where the $|_{m=m_n}$ indicates that the Bessel and Neumann functions are to be evaluated at m = $m_n$.
One can then convert the sum over $m_n$ to an integral
\begin{equation}
\sum_{m_n} \rightarrow \int dm_n \frac{e^{\beta y_2}}{\beta\pi}.
\end{equation}
when the poles become dense as in the examples of Eqs.(52,53).

\section{\it{Newtonian potential}}

In the static approximation, 1/$m^2$ $\rightarrow$ -1/$\vec{p}^2$.  After a Fourier transformation, the 1/$m^2$ part of $h_{00}(\vec{p},y_{\alpha})$ then gives rise to the 1/r Newtonian potential in the static limit.  Since $h_{00}$ = $h_{00}^{TT}$ - $f^{T}$/3 we can combine Eq.(48) and Eq.(55) to obtain for the 1/r part on the $y_1$ brane
\begin{equation}
h_{00}(y_1) = -\frac{4\beta}{3M_{5}^{3}m^2}[T_{00}(y_1) + e^{-2\beta y_2}T_{00}(y_2)] + \frac{\beta}{3M_{5}^{3}m^2}T_{00}(y_1).
\end{equation}
where the first term is $h_{00}^{TT}$ and the second is -$f^T$/3.  Combining terms one gets
\begin{equation}
h_{00}(y_1) = -\frac{\beta}{M_{5}^{3}m^2}T_{00}(y_1) - \frac{4\beta e^ {-2\beta y_2}}{3M_{5}^{3}m^2}T_{00}(y_2)
\end{equation}
The stress tensor for a point particle is
\begin{equation}
T^{\mu\nu} = \frac{1}{\sqrt{-g}}\frac{\delta\mathcal{L}_{matter}}{\delta g_{\mu\nu}} = e^{4\beta y_2}\frac{\delta\mathcal{L}_{matter}}{\delta g_{\mu\nu}}
\end{equation}
and hence $T_{00}(y)$ = $e^{2\beta y}\bar{m}\delta^{3}(x^i - x'^i)$ where $\bar{m}$ = $m_{o}e^{-\beta y}$ is the mass of a particle on the brane at position y.  One then finds for the interaction energy between two particles
\begin{equation}
V(y_1 = 0) = -\frac{\beta\bar{m}_1\bar{m}'_1}{2M_{5}^{3}m^2} - \frac{2\beta\bar{m}_1\bar{m}_2}{3M_{5}^{3}m^2}
\end{equation}
where $\bar{m}_1, \bar{m}'_1$ are masses on the $y_1$ = 0 brane and $\bar{m}_2$ = $m_{0}e^{-\beta y_2}$ is a mass on the $y_2$ brane.  We can then read off Newton's constant for each term
\begin{displaymath}
G_N = \left\{ \begin{array}{ll}
\frac{\beta}{8\pi M_{5}^{3}} & \textrm{between matter on the $y_1$ = 0 brane}\\
\frac{4}{3}(\frac{\beta}{8\pi M_{5}^{3}}) & \textrm{between matter at $y_1$ = 0 and $y_2$}
\end{array}\right.
\end{displaymath}
We see that the Newtonian constant depends on whether both particles are on the $y_1$ = 0 (Planck) brane or one is on the $y_2$ (TeV) brane.  The extra 4/3 factor for the $y_2$ particle arises because the $f^T$ part of Eq.(58) has no $T_{00}(y_2)$ term.  We can similarly calculate $h_{00}(y_2)$ which then gives the potential at $y_2$ to be
\begin{equation}
V(y_2) = -\frac{2\beta\bar{m}_2\bar{m}'_2}{3M_{5}^{3}m^2} - \frac{\beta\bar{m}_2\bar{m}'_2}{6M_{5}^{3}m^2}e^{2\beta y_2} - \frac{2\beta\bar{m}_1\bar{m}'_2}{3M_{5}^{3}m^2}
\end{equation}
where $\bar{m}_{2}$ = $m_{20}e^{-\beta y_2}$, $\bar{m}'_{2}$ = $m'_{20}e^{-\beta y_2}$ are masses on the $y_2$ brane, and $\bar{m}_1$ = $m_{10}$ on the $y_1$ = 0 brane. (Here the subscript 0 indicates the mass before it is rescaled by the exponential factor.)  In $V(y_2)$ the $e^{2\beta y_2}$ in the second term (arising from -$f^T$/3) is not absorbed in rescaling the masses on the TeV ($y_2$) brane in the second term which means that this term will dominate the potential giving an enormous attractive force that differs from the standard Newtonian form.

\section{\it{Conclusions}}

We have examined gravitational forces in a 5D model with two 3-branes and $S^1$/$Z_2$ symmetry.  The coordinate conditions we use, $h_{5i}$ = 0, produce no brane bending effects.  While a stabilizing field has not been included in the analysis, the analysis is still consistent since $h_{55}$ $\neq$ 0 and so the branes are not constrained to be fixed. (All the Einstein equations are seen to give consistent solutions.)

The static Newtonian potential comes from two parts of the 4D piece of the metric: $h_{ij}^{TT}$ and $f^T$.  While $h_{ij}^{TT}$ sees matter on both branes, $f^T$ at $y_{\alpha}$ sees matter only on the $y_{\alpha}$ brane.  In RS, the $f^T$ contribution on the Planck brane is what is needed to give the correct relation between the 5D Planck mass, $M_5$, and the Newton constant $G_N$, i.e.
\begin{equation}
G_N = \frac{\beta}{8\pi M_{5}^{3}}
\end{equation} 
for forces between Planck brane particles.  However, on the TeV brane, even after the masses are appropriately rescaled, the $f^T$ contribution on the TeV brane gives an attractive force proportional to $e^{2\beta y_2}$ $\gg$ 1.  It remains to be seen if this problem can be eliminated by including a stabilizing mechanism, e.g., a Goldberger-Wise scalar field in the bulk which modifies the vacuum metric. This would then modify the perturbation to the metric $h_{00}$ produced by matter on the branes.

\section{\it{Acknowledgements}}

This work was supported in part by a National Foundation Grant Phy-0101015.

\end{document}